\documentstyle[12pt,psfig]{article}

\newcommand{\resection}[1]{\setcounter{equation}{0}\section{#1}}
\newcommand{\appsection}{\addtocounter{section}{1} \setcounter{equation}{0}
                         \section*{Appendix \Alph{section}}}

\def\de {\mbox{d}}
\def\be{\begin{equation}}
\def\ee{\end{equation}}
\def\bd{\begin{displaystyle}}
\def\ed{\end{displaystyle}}
\def\ba{\begin{array}}
\def\st{\stackrel}
\def\ea{\end{array}}
\def\EQ{\begin{equation}}
\def\EN{\end{equation}}
\def\bea{\begin{eqnarray}}
\def\eea{\end{eqnarray}}
\def\beano{\begin{eqnarray*}}
\def\eeano{\end{eqnarray*}}

\def\th{\theta}
\begin{document}
\oddsidemargin 5mm
\setcounter{page}{0}
\newpage     
\setcounter{page}{0}
\begin{titlepage}
\begin{flushright}
ISAS/EP/96/105

IC/96/168 

\end{flushright}
\vspace{0.5cm}
\begin{center}
{\large {\bf On the Form Factors of Relevant Operators \\
and their Cluster Property}}\\
\vspace{1.5cm}
{\bf C. Acerbi$^{a,b}$, G. Mussardo$^{a,b,c}$ and A. Valleriani$^{a,b}$
}\\
\vspace{0.5cm}
{\em $^a$ International School for Advanced Studies, \\
Via Beirut 3, 34014 Trieste, Italy} \\
\vspace{1mm} 
{\em $^b$ Istituto Nazionale di Fisica Nucleare, \\
Sezione di Trieste, Italy} \\
\vspace{1mm}
{\em $^c$ International Centre of Theoretical Physics,\\
Strada Costiera 11, 34014 Trieste, Italy} \\
\end{center}
\vspace{3mm}
\begin{abstract}
\noindent
We compute the Form Factors of the relevant scaling operators in a class of 
integrable models without internal symmetries by exploiting their cluster 
properties. Their identification is established by computing the 
corresponding anomalous dimensions by means of Delfino--Simonetti--Cardy 
sum--rule and further confirmed by comparing some universal ratios of the 
nearby non--integrable quantum field theories with their independent 
numerical determination.  
\end{abstract}
\vspace{5mm}
\end{titlepage}
\newpage
\setcounter{footnote}{0}

\resection{Introduction}
In this work we present a detailed investigation of the matrix elements 
\EQ
F^{\cal O}_{a_1,a_2,\ldots,a_n}(\beta_1,\ldots,\beta_n) \,=\,
\langle 0 \mid {\cal O}(0)\mid A_{a_1}(\beta_1) \ldots A_{a_n}(\beta_n)\rangle 
\label{FF}
\EN
(the so--called Form Factors (FF)) in a class of integrable two--dimensional 
quantum field theories. Our specific aim is to check some new theoretical 
ideas which concern the relationships between three different regimes which 
two--dimensional quantum field theories may have, namely the ones ruled by 
conformal invariance, integrable or non--integrable dynamics. 

Conformal Field Theories (CFT) and the associated off-critical 
Integrable Models (IM) have been extensively studied in the last years: 
as a result of these analyses a good deal of information has been obtained 
particularly on correlation functions of a large number of statistical 
mechanical models in their scaling limit and on physical quantities 
related to them (see for instance [1-8]). In this context, a crucial 
problem often consists in the determination of the spectrum of the 
scaling operators away from criticality, namely their correct identification 
by means of the set of their Form Factors. This is one of the issues 
addressed in this work. 

Form Factors also play a crucial role in estimating non--integrable 
effects. Let us first recall that the above CFT and IM regimes cannot
obviously exhaust all possible behaviours that statistical models 
and quantum field 
theories can have since typically they do not possess an infinite 
number of conservation laws. This means that in general we have to face 
all kinds of phenomena and complications associated to Non--Integrable Models 
(NIM). The scattering amplitudes and the matrix elements of the quantum 
operators will have in these cases a pattern of analytic singularities due 
both to the presence of higher thresholds and to the appearance of 
resonances. A first step forward in their analysis has been recently 
taken in ref.\,\cite{NIM} where it has been shown that some interesting 
examples of NIM may be obtained as deformations of integrable models. 
The action of such theories can correspondingly be written as        
\EQ
{\cal A} \,=\,{\cal A}_{int} + 
\sum_i\,\lambda_i\int d^{\,2} x \, \Psi_i(x) \,\,\,,
\label{action}
\EN
${\cal A}_{int}$ being the action of the integrable model. 
Since the exact expressions (\ref{FF}) of the Form Factors of the integrable 
theories are all assumed calculable, in particular the ones of the fields 
$\Psi_i(x)$ entering eq.\,(\ref{action}), one is inclined to study the 
non--integrable effects by using the Born series based on the Form Factors. 
Although at first sight this still remain a difficult task (and generally, 
it is indeed so), there may be favorable circumstances where the analysis 
simplifies considerably. For instance, as far as there is only a soft 
breaking of integrability, it has been shown in \cite{NIM} that 
the complications of the higher terms in the series can often be avoided   
since the most relevant corrections only come from the lowest approximation. 
If this is the case, one can  extract an important amount of 
information with relatively little effort: a significant set of physical 
quantities to look at is provided for instance by universal ratios, like 
the ones relative to the variations of the masses or of the vacuum energy 
density ${\cal E}_{vac}$: if the breaking of integrability is realized 
by means of a single field $\Psi(x)$, those are expressed by 
\be
\ba{l}
\bd
 \frac{\delta m_i}{\delta m_j} \,=\, \frac{m_j^{(0)}}{m_i^{(0)}} 
\frac{F_{ii}^{\Psi}(i \pi)}{F_{jj}^{\Psi}(i \pi)} \,\, , \ed  \\  
\\
\bd \frac{\delta {\cal E}_{vac}}{m_1^{(0)} \delta m_1} \, = \,  
\frac{\langle 0 \mid \Psi\mid 0\rangle}{F_{11}^{\Psi}(i \pi)} \,\, ,  
\ed
\ea 
\label{nif}
\ee
where $m_i^{(0)}$ refers to the (unperturbed) mass spectrum of the 
original integrable theory. It is thus evident that also to  
estimate the non--integrable effects associated to a given operator $\Psi(x)$
one  must face the problem of correctly identifying its FF's.   

Two new results on the relationship between CFT and IM have been recently 
derived by Delfino, Simonetti and Cardy \cite{DSC}. Briefly stated, the 
first result consists in a new sum--rule which relates the conformal dimension 
$\Delta^{\phi}$ of the operator $\phi(x)$ to the off--critical (connected) 
correlator $\langle \Theta(x) \phi(0) \rangle_c$, where $\Theta(x)$ is the 
trace of the stress--energy tensor\footnote{The sum--rule 
in the form of eq.\,(\ref{sumrule1}) may be violated by effect of 
renormalization of the operators outside the critical point, as 
clarified in the original reference \cite{DSC}. This is however 
not the case for the field theories and the operators considered 
in this work.}  
\EQ
\Delta^{\phi} \,=\, -\frac{1}{4 \pi \langle \phi\rangle} 
\int d^2x \, \langle \Theta(x) \phi(0) \rangle_c \,\,\, .
\label{sumrule1}
\EN
This sum--rule is closely related to the analogous expression for the 
conformal central charge $c$ \cite{ctheorem}
\EQ
c\,=\, \frac{3}{4 \pi} 
\int d^2x \, \mid x\mid^2 \langle \Theta(x) \Theta(0) \rangle_c \,\,\, .
\label{sumrule2}
\EN
Equations (\ref{sumrule1})  and (\ref{sumrule2})  express elegant relationships between conformal 
and off--critical data, but more importantly, they provide very concrete and 
efficient tools to characterise the scaling limit of the off-critical models. 

As for the second result, it has been suggested by the aforementioned 
authors of ref. \cite{DSC}, that the Form Factors of the relevant scaling 
fields\footnote{Hereafter we are using the short expression ``scaling 
fields" to actually denote the off-critical operators which reduce to 
the scaling fields in the conformal limit.} of an integrable field theory 
-- in absence of internal symmetries -- are in one--to--one correspondence 
with the independent solutions of the so called {\em cluster equations} 
\bea  
\label{cluster}
\lim_{\Lambda \rightarrow \infty} 
       F^\Phi_{a_1,a_2,\ldots,a_k,a_{k+1},\ldots,a_{k+l}} 
       (\th_1,\th_2,\ldots,\th_k,\Lambda + \th_{k+1},
       \ldots,\Lambda + \th_{k+l}) &=& 
        \\
      & &\!\!\!\!\!\!\!\!\!\!\!\!\!\!\!\!\!\!\!\!\!\!\!\!\!\!\!\!\!\!\!
         \!\!\!\!\!\!\!\!\!\!\!\!\!\!\!\!\!\!\!\!\!\!\!\!\!\!\!\!\!\!\!
         \!\!\!\!\!\!\!\!\!\!\!\!\!\! \!\!\!\!\!\!\!\! \!\!\!\!\!\!\!\! 
           = \frac{1}{\langle \Phi\rangle} 
       F^\Phi_{a_1,a_2,\ldots,a_k}(\th_1,\th_2,\ldots,\th_k)\,
         F^\Phi_{a_{k+1},\ldots,a_{k+l}}(\th_{k+1},\ldots,\th_{k+l})
         \nonumber \, 
\eea 
These equations can be imposed on the Form Factors {\em in addition} to the 
customary functional and residue equations which they satisfy (see in this 
respect also \cite{Smirnov,SG}). If this {\em cluster hypothesis} 
is valid, we would have 
 a clear method to identify the matrix elements of all the 
relevant operators, at least in the case of theories without 
symmetries. It must be stressed that until now this task has often 
been a matter of keen guess--work and mostly based on physical intuition. 

It turns out that a check of the above {\em cluster hypothesis} provides
a well--suited forum for testing several theoretical aspects. In fact, the
most  direct way of confirming the above idea is firstly to solve the 
general functional equations of the Form Factors with the additional 
constraints of the cluster equations (\ref{cluster}) and to see whether  
{\em the number of independent solutions equals the number of relevant 
fields} in the corresponding Kac table. If the above check turns out to 
be positive, one may use the sum--rule (\ref{sumrule1}) in order 
to achieve the correct identification of the (supposed) primary relevant 
operators $\phi_i$: from the values of the partial sums one can in fact 
infer the value of the anomalous dimension and correspondingly 
recognize the operator. An additional confirmation 
may also come from the employment of eqs.\,(\ref{nif}) 
relative to non--integrable field theories. In fact, one can regard the 
primary field $\phi_i(x)$ under investigation as that  
operator which spoils the integrability of the original theory and  
therefore compare the predictions (\ref{nif}) based on its Form Factors 
with their independent numerical determinations which may be obtained by 
means of the truncation method \cite{TCS}. Note that a successful 
test of this kind could also be interpreted the other way around, namely as 
a further proof of the effectiveness of the formulas (\ref{nif}) in 
estimating  non--integrable effects. 

The models on which we have chosen to test the above considerations 
are integrable deformations of the first representatives of the  
non--unitary
conformal  series\footnote{The conformal weigths and central charge
are given respectively by 
\bea 
&& \Delta_{1,a} = - \frac{(a - 1)\, (2n -a)}{2 \,(2n +1)}, \hspace{3mm} 
a = 1,2,\ldots, 2n \nonumber \\ && \nonumber \\
&& c = -\frac{2 \,(6 n^2 - 7 n +1)}{2 n+1} \,\, , \hspace{5mm} n=2,3,\ldots 
\nonumber 
\eea
}
${\cal M}(2,2\,n+1)$, $n \geq 2$.
They belong to the class of universality of solvable RSOS lattice models 
{\em \`{a} la} Andrews--Baxter--Forrester although with negative Boltzmann 
weigths \cite{ABF,Riggs}: their simplest example is given by the quantum 
field theory associated to the so--called Yang--Lee model which describes 
the distribution of zeros in the grand canonical partition function of the 
Ising model in a complex magnetic field \cite{Fisher,CMYL}. These models 
do not have any internal symmetry and all their fields are relevant 
operators: hence, they are ideal  for our purposes. Moreover, 
the nature of their massive and conformal phases is simple enough. The
price to pay for their relative simplicity is however the presence of 
typical non--unitary phenomena, as imaginary coupling constants or negative 
values of the anomalous dimensions and central charge, together 
with the anomalous poles in the $S$--matrix which induce an 
unusual analytic structure in the Form Factors \cite{BPZ,CMYL,KMM}. 

The paper is organized as follows. In Section 2 we discuss the general 
strategy which can be employed in order to  compute the FF's 
of the relevant operators in the integrable deformations of the models 
${\cal M}(2,2\,n+1)$. In Section 3 and 4 we present a detailed analysis 
of the FF's of the models ${\cal M}(2,7)$ and ${\cal M}(2,9)$, which are 
the first non--trivial examples on which to check all the theoretical 
ideas discussed above. In fact, for the first model ${\cal M}(2,5)$ 
the {\em cluster hypothesis} is easily verified: the only solution of the 
Form Factor equations is the sequence of functions determined in ref. \cite{YL} 
which indeed fulfill the cluster equations (\ref{cluster}) and are easily 
identified with the matrix elements of the only relevant field 
of the Yang--Lee model. The two models ${\cal M}(2,7)$ and ${\cal M}(2,9)$
represent, somehow, the best playground for our purposes 
because they give rise to integrable models under each of their 
possible (individual) deformations and also because they optimize 
the size of the lenghty numerical output which we present for the 
solutions of the non--linear equations. Moreover, although there 
is in principle no obstacle to extend the analysis to all the models 
${\cal M}(2,2\,n+1)$, these are the simplest cases from a computational 
point of view since the larger is the value of the index $n$ 
the higher is the order of the system of algebraic equations to be 
solved for determining the Form Factors. Finally, our conclusions 
are in Section 5. Two appendices complete the paper: Appendix A 
gathers all important formulas relative to the parameterization 
of the two--particle Form Factors and Appendix B collects the 
$S$--matrices of the models analysed.  

\resection{Outline of Our Strategy}

In this section we discuss the general strategy needed in order to obtain 
the Form Factors of the scaling primary fields of the integrable 
deformations $\phi_{1,k}$ of the conformal models ${\cal M}(2,2n+1)$ 
(hereafter denoted by the shorthand notation $[{\cal M}(2,2n+1)]_{(1,k)}$).
The deforming field $\phi_{1,k}$ can be one of the operators $\phi_{1,2}$,  
$\phi_{1,3}$ or possibly some other primary field which gives rise to an 
integrable deformation. 

The starting point in the computation of the Form Factors is 
the correct parameterization of the two--particle ones which is given 
detailed in Appendix A. This is a non--trivial task in the case of 
non--unitary models for the reason that the exact $S$--matrices of 
these models are usually plagued by a pletora of anomalous poles 
\cite{KMM}. By this we mean for example simple poles which are not 
related to any bound state, or, more generally, any poles which do not 
have apparently the standard diagrammatic interpretation of 
refs.\,\cite{multiple}. Consider for example the $S$--matrices 
listed in the tables of Appendix B relative 
to the integrable deformations of the models ${\cal M}(2,7)$ and 
${\cal M}(2,9)$ where the anomalous poles have been labelled 
with ${\cal B}$, ${\cal D}$ or $*$. The origin of these poles 
may be explained according to the ideas put forward in \cite{generalized}. 
In particular, poles of type ${\cal B}$ and ${\cal D}$ 
are due to multiparticle processes of the kind described respectively 
by the ``butterfly'' and ``dragonfly'' diagrams drawn in 
Figures 2 and  3 respectively. These multi--loop processes 
induce in the $S$--matrix simple poles rather than higher order 
ones because the internal lines of these diagrams cross at relative 
rapidity values relative to some zeros of their corresponding 
two--particle $S$--matrix element: this gives rise to a partial 
cancellation of the poles.

The adopted parameterization for two--particle FF's is directly related 
to the pole structure of the $S$--matrix. This yields to the 
expression (\ref{Fab}) whose functional form is set except for the 
coefficients $a_{ab,\Phi}^{(k)}$ appearing in the expansion (\ref{qgen}) 
of the polynomials $Q_{ab}^{\Phi}(\th)$. The degree $k_{ab,\Phi}^{\rm max}$ 
of these polynomials is fixed by the asymptotic behavior of the FF's 
for large rapidities which depends, of course, on the field $\Phi$ \cite{DM}. 
For the case of two--particle FF's of cluster operators, it is easy to 
see that they are subject to have for large $\th$ at most a constant 
limit\footnote{The limit may vanish in the presence of symmetries.}. 
In fact, for two--particle FF's eqs. (\ref{cluster}) read 
\be  \label{cluster2} 
\lim_{\th \rightarrow \infty} F_{ab}^\Phi(\th) = 
F_a^\Phi  \, F_b^\Phi\,\,. 
\ee
Hereafter we deal with dimensionless cluster operators 
which are normalized in such a way as to have a vacuum expectation value 
equal to one\footnote{Since the relevant primary operators will be identified 
with the cluster ones 
(except their dimensional factors which can be easily restored), 
in the sequel we will adopt the same normalization also for them.}    
\be 
     \langle 0 | \Phi (0) | 0 \rangle = F_0^\Phi = 1 \,\,.
\ee
In order to fully determine the FF's of the cluster operators  
we have chosen to focus on the set of all one-- and two--particle FF's.
Listing all the relations among them, one obtains a system 
of equations in the unknown parameters $F_a^\Phi$ and $a_{ab,\Phi}^{(k)}$.
Let us see then all information we have on the FF's. 
 
The first equations that  one must consider are the {\em dynamical residue 
equations} resulting from the detailed analysis of the poles they are 
endowed with. These equations relate FF's with different external particles 
and may have a different origin. In particular, for every simple bound state 
pole of the amplitude $S_{ab}$ at angle $\th = i u_{ab}^{c}$ relative 
to the particle $A_c$ (see Figure 1), we have  
\be
\label{boundfpole}
\lim_{\th \rightarrow i u_{ab}^{c}}(\th -iu_{ab}^{c})
F^{\Phi}_{ab}(\th)= i \,\Gamma_{ab}^{c} F^{\Phi}_{c} \,\, ,
\ee
where the on--mass--shell three--point coupling constant $\Gamma_{ab}^c$ 
is given by the residue on the pole of the $S$--matrix
\begin{equation}
\label{gamma}
 \lim_{\theta\rightarrow i u_{ab}^{c}}
(\theta - iu_{ab}^{c})
S_{ab}(\theta)= i\, (\Gamma_{ab}^{c})^2 \,\,\,.
\end{equation}
Dynamical residue equations are also provided by double order poles and 
simple order poles of type ${\cal B}$. Both of them are related  
to diagrams of the kind shown in Figure 2. For each such diagram, 
one can write the following equation
\be
\label{doubfpole}
  \lim_{\th_{ab}\rightarrow i\varphi}(\th_{ab}-i\varphi) 
F_{ab}^{\Phi}(\th_{ab}) \,=\,i\,
\Gamma_{ad}^c \,\Gamma_{db}^e \,F^{\Phi}_{ce}(i\gamma)\,\, ,
\ee
where $\gamma=\pi - u_{cd}^{a}- u_{de}^{b}$. In the case of ${\cal B}$ poles  
one can always verify that the amplitude $S_{ce}(\th)$ has a simple zero at 
$\th = i\gamma$. More complicated residue equations can be in general obtained 
with reference to ${\cal D}$ poles and higher order ones whose explicit 
expressions -- not reported here -- can be  however easily written, once 
the corresponding multi--scattering diagrams have been identified. 

It must be stressed that the above set of equations just depend on the 
dynamics of the model through its $S$--matrix and hold identical for 
every operator $\Phi (x)$. Therefore, in general, some residual freedom 
on the parameters is still expected after imposing these equations, 
because they must be satisfied by the FF's of all operators compatible 
with the assumed asymptotic behaviour.

Adding to this system of {\em linear} equations the {\em non--linear} 
cluster equations (\ref{cluster2}) of the two--particle FF's, one obtains
in general a redundant set of compatible equations in all the unknown 
parameters of the one-- and two--particle FF's. Due to its non--linearity,
the system allows a multiplicity of solutions which define the 
so--called {\em cluster operators} of the theory\footnote{In all cases 
analyzed, the smallest system of equations among different FF's which 
is sufficient to determine their coefficients turns out to involve just 
a subset of the two--particle FF's. This suggests that also in the general 
case it should be possible to predict the final number of cluster 
solutions already from a ``minimal'' system, avoiding in this way to deal 
with systems of equations involving a huge number of unknown variables.}. 
If the number of solutions of the system matches the cardinality 
of the Kac table of the model one is led to identify them with the families 
of FF's of the relevant primaries.

Among the cluster solutions, one can first of all identify the FF's 
of the deforming field $\phi_{1,k}$. This operator is known to 
be essentially the trace of the energy--momentum tensor $\Theta(x)$ 
since 
\be \label{Th}
\Theta (x) = 4 \pi\, {\cal E}_{vac} \, \phi_{1,k} \,\, ,
\ee
${\cal E}_{vac}$ being the vacuum energy density which
can be easily computed by TBA computations \cite{TBA}
\be
{\cal E}_{vac} = - \frac{m_1^{\,2}}{8 \,\sum_{x \in P_{11}} \sin (\pi x)}
\,\,\, .
\ee
Here the set $P_{11}$ is defined in eq.\,(\ref{Sab}) and $m_1$ is the 
lightest particle mass. In view of the proportionality (\ref{Th}), the 
FF's of $\phi_{1,k}$ can be identified among the cluster solutions by 
checking the peculiar equations which characterize the two--particle 
FF's of $\Theta (x)$ in virtue of the conservation of the energy--momentum 
tensor, namely the normalization of the diagonal two--particle FF's 
\be 
\label{FThetadiag}
         F^\Theta_{aa}(i \pi) = 2 \pi m_a^2 \:,
\ee
and the factorization of the polynomial $Q^{\Theta}_{ab}$ 
for non--diagonal two--particle FF's ($a \neq b$) into 
\be 
\label{FThetanondiag}
Q^{\Theta}_{ab} (\cosh \th)  \,= \, 
\left( 2\,m_a \,m_b \:\cosh \th + m_a^2 +  
m_b^2 \right) \: R^{\Theta}_{ab}(\cosh\th)\, ,  
\ee
where $R^{\Theta}_{ab}$ is  a suitable polynomial \cite{Smirnov,DM}. 

Knowing the FF's of $\Theta (x)$, one is then enabled to make use of 
the sum--rule (\ref{sumrule1}) to compute  the conformal dimension 
of the operators defined by the remaining cluster solutions in order
to identify them with all the relevant primaries of the theory.
This sum--rule can be evaluated by using the spectral representation 
of the correlator  
\be
\label{formexp}
\langle\,\Theta(x) \phi(0)\,\rangle_c\, =
\sum_{n=1}^{\infty} \sum_{a_i}\int_{\th_1>\th_2\ldots>\th_n} 
\frac{\de^n\th}
{(2\pi)^n}\, F_{a_1,\ldots,a_n}^{\Theta}
({\bf\theta}) \,F_{a_1,\ldots,a_n}^{\phi}
(i\pi-\th)\,
e^{-|x|\sum_{k=1}^{n}m_k\cosh\theta_k} \,\,\, .
\ee
In all the models we have studied, the corresponding series for the 
sum--rule (\ref{sumrule1}) displays a very fast convergence behaviour 
for any of the cluster operators. The truncated sums obtained by 
including just very few contributions have proved sufficient to attain 
a good approximation of all the values expected by the Kac table of 
conformal dimensions. In this way, the one--to--one correspondence 
between cluster solutions and primary relevant operators can been 
easily set. 

Finally, having obtained the FF's of all the relevant fields in each 
integrable deformation, as a further check of their correct identification, 
one may employ the formulas (\ref{nif}) relative to the universal ratios  
of the nearby non--integrable quantum field theories. These predictions 
can then be compared against their numerical estimates obtained from 
Truncated Conformal Space (TCS) approach developed in \cite{TCS}. 
The agreement between numerical estimates and theoretical predictions 
of the non--integrable effects may provide additional confirmation 
and may remove all possible remaining doubts about the validity of 
the cluster hypothesis for these models.

\resection{Integrable Deformations of ${\cal M}{(2,7)}$}

The minimal conformal model ${\cal M}{(2,7)}$ has, in addition to 
the identity operator $\phi_{1,1}$, only two primary operators, 
$\phi_{1,2}$ and $\phi_{1,3}$, both of them relevant with conformal 
weights given by $-2/7$ and $-3/7$ respectively \cite{BPZ}.
The perturbations of the conformal action either by the 
``magnetic operator'' $\phi_{1,2}$ or by the ``thermal operator'' 
$\phi_{1,3}$ are both known to be, separately, integrable 
\cite{KMM}. The $S$--matrices and the mass ratios of the 
two integrable models are given in tables B1 and B2.
In their massive phase, both perturbations have two stable 
massive particles denoted by $A_1$ and $A_2$, with a mass 
ratio and a scattering matrix which depend on the integrable 
direction considered. In each case, we expect to find two 
non--trivial independent families of Form Factors solutions to 
the cluster equations (\ref{cluster}) (in addition to the family of 
the null Form Factors relative to the identity operator).

The Form Factors of the primary operators of the model relative 
to the thermal deformation have already been considered in 
\cite{koubek}. Here, we have performed an {\em ab--initio} calculation  
by imposing the cluster equations: our result has been in perfect 
agreement with the FF's of ref.\,\cite{koubek}, proving in this way that 
these cluster solutions are also unique.

The result of the computation of Form Factors in the 
two integrable deformations  $[{\cal M}(2,7)]_{(1,2)}$  
and $[{\cal M}(2,7)]_{(1,3)}$ are summarised in tables 
1--2 and 3--4 respectively where we list the values of 
the one--particle FF's and the coefficients $a_{ab,\phi}^{(k)}$ 
of the two--particle FF's relative to some of the lightest 
two--particle states. As expected, we find two non--trivial 
solutions of Form Factors families. In each deformation, the FF's 
of the deforming operator suitably rescaled by (\ref{Th}), 
can be immediately identified because they satisfy 
the peculiar equations characterizing the trace of the 
energy--momentum tensor (\ref{FThetadiag}) and (\ref{FThetanondiag}).
This is further confirmed by employing the spectral representation 
of the correlator $\langle\Theta (x) \Theta (0)\rangle_c$ in 
the sum--rule (\ref{sumrule2}), which provides in both deformations 
the value of the central charge with a very high precision 
(the relative error being of order $10^{-4}$--$10^{-5}$). 
The identification of both the solutions with the primaries 
$\phi_{1,2}$ and $\phi_{1,3}$ is easily established after computing for 
each solution its UV anomalous dimension by means of the sum 
rule (\ref{sumrule1}). The contributions to this sum rule coming from the 
dominant lightest multiparticle states are given in 
Tables 5 and 6 for the two deformations (the contributions 
are ordered according to increasing values of the Mandelstam variable 
$s$ of the multi--particle state). The agreement of the truncated sums 
with the known values of the anomalous dimensions is very satisfactory 
given the fast convergency behaviour of the spectral series. 
In the computation of these sum rules, some three--particle FF 
contributions have been inserted as well, although we do give here 
their exact expression for sake of simplicity (their general 
parameterization follows the one adopted for instance,
in \cite{DM}). It should be noticed that the oscillating 
behaviour of these sums is typical of non--unitary theories 
where one expects, in general, both positive and negative terms. 

\subsection{Non--Integrable Deformations of ${\cal M}(2,7)$}

For each possible integrable deformation of the model, the addition of 
a further orthogonal deformation breaks its integrability leading, 
among other things, to corrections of the mass spectrum and of 
the vacuum energy. Both corrections can be independently computed 
by performing a numerical diagonalization of the off--critical 
Hamiltonian by means of the so--called Truncation Method \cite{TCS}. 
We have carried out this analysis comparing these non--integrable data
with the theoretical predictions by eqs.\,(\ref{nif}). Let us briefly 
describe the output of these studies. 

The double non--integrable deformation
\[
[{\cal M}(2,7)]_{(1,3)} + \varepsilon  \phi_{1,2}\, ,
\]
for small values of $\varepsilon m_1^{2\Delta_{1,2} -2}$ has 
already been studied in \cite{NIM}, where a good agreement between 
numerical and theoretical values has been found. Having obtained the 
FF's for the $\phi_{1,2}$ deformation, we are now able to complete 
the analysis by testing the opposite deformation
\[
[{\cal M}(2,7)]_{(1,2)} + \varepsilon  \phi_{1,3}\,\,\,.
\]
The numerical determination of the two universal ratios of eq.\,(\ref{nif}) 
(for small values of $\varepsilon m_1^{2\Delta_{1,3} -2}$) 
gives $\frac{\delta m_1}{\delta m_2}= 0.675$ and 
$\frac{\delta {\cal E}_{vac}}{\delta m_1}= -0.244 \, m_1^{(0)}$
with a precision estimated to be up to a few percents. This values fully 
agree with the computed theoretical values $\frac{\delta m_1}{\delta m_2} 
= 0.68404$ and $\frac{\delta {\cal E}_{vac}}{\delta m_1} = -0.24365 \, 
m_1^{(0)}$ (see, for instance Figure 4 and 5 where the data relative to 
the ratios $\frac{\delta m_1}{\delta m_2}$ and 
$\frac{\delta {\cal E}_{vac}}{\delta m_1}$ respectively
are reported for different values of $\varepsilon$).

\resection{Integrable Deformations of ${\cal M}(2,9)$}

In this section, we turn our attention to the ${\cal M}{(2,9)}$ minimal 
model which displays a richer structure in the RG space of relevant couplings.
This model has in fact, besides the identity, three primary operators 
$\phi_{1,2}$, $\phi_{1,3}$ and $\phi_{1,4}$ which are all relevant with 
conformal dimensions $-1/3$, $-5/9$  and $-2/3$ respectively. These fields 
taken separately give rise to different integrable deformations of the 
conformal model, each of them characterized by a different mass spectrum and 
$S$--matrix (see tables B3, B4 and B5 in Appendix B). In particular, the 
first two deformations produce three--particle mass spectra (with different 
mass ratios) while the last one gives a four--particle spectrum. 

The FF's of the primary operators in the $\phi_{1,3}$--deformation had 
already been obtained in ref.\,\cite{koubek} and were known to satisfy 
the cluster property. Again, our derivation of these FF's as solutions 
of the cluster equations proves that the FF's found in \cite{koubek} 
are the only possible cluster solutions.

The Form Factors of the cluster solutions for each of the three 
above mentioned deformations have been computed according to the 
strategy explained in Section 2. The resulting one--particle FF's and 
two--particle FF's coefficients are given in tables 7--8, 9--10 and 
11--12 respectively. The important result is that in each integrable 
deformation of this model, three families of non--trivial solutions 
have been found. Among the solutions, we have firstly identified the FF's of 
the deforming field by checking the exact fulfillment of 
eqs.\,(\ref{FThetadiag}) and (\ref{FThetanondiag}), after the 
appropriate rescaling (\ref{Th}). Moreover, the $c$--sum--rule 
(\ref{sumrule2}) can be easily shown to give very precise approximations 
of the central charge in each of the three separate deformations.

As for the other solutions, they have been successfully identified 
with the FF's of the primary operators by computing their anomalous 
dimension by means of eq. (\ref{sumrule1}). The first contributions 
to these sums are given in tables 13, 14  and 15. In all cases the 
agreement with the expected anomalous dimensions of the primaries is 
established, even though the convergence of the series is observed 
to be noticeably faster for lower absolute values of the anomalous 
dimension of the deforming field. This observed trend is indeed expected 
from the short--distance behavior of the correlator (\ref{formexp}), 
as predicted by the Operator Product Expansion of the fields. In fact, 
in the models ${\cal M}{(2,2\,n+1)}$ where the fields have negative 
anomalous dimensions, this correlator displays a zero at the origin 
whose power law exponent is larger for lower absolute values of the 
anomalous dimension of $\Theta(x)$; correspondingly, the small $x$ region of 
integration in (\ref{sumrule1}) is less relevant making the lightest  
multiparticle states more dominant in the series.

\subsection{Non--Integrable Deformations of ${\cal M}(2,9)$}

The availability of the FF's of all the primary fields of the model 
has allowed us, in each of the three separate integrable deformations, 
to consider two different orthogonal non--integrable deformations. 
We have had then the possibility of testing the theoretical values 
obtained for the universal quantities (\ref{nif}) versus their numerical 
TCS estimates in six different multiple deformations, exploring in this 
way the non--integrable region around the conformal point of the model.
The outcome of the analysis in all the deformations is summarized in 
Table 16. Since the precision of TCS data is expected to be of approximately 
a few percents, the comparison with the computed theoretical values is in 
all cases quite satisfactory.

\resection{Conclusions}

The main purpose of this work has been to substantiate by means of concrete 
{\em ab--initio} calculations the cluster hypothesis for the Form Factors 
of the relevant operators in integrable quantum field theories obtained as
deformation of a conformal action. We have studied, in particular, the 
matrix elements of the primary operators in the integrable deformations 
of the first models of the non--unitary series ${\cal M}(2,2n+1)$. 
In all cases analysed, we have confirmed the cluster hypothesis since 
we have found a one--to--one correspondence between the independent 
solutions of the cluster equations and the relevant fields. 

It should be said that the absence of internal symmetries of the above models 
has played an important role in carrying out our computations. In fact, in 
this situation one can exploit the cluster equations (\ref{cluster}) in their 
full generality. It would be interesting to see how the results of 
ref.\,\cite{DSC} generalize to the case of quantum field theories with 
internal symmetries which induce selection rules on the matrix 
elements. Another important open problem is also to understand the meaning of 
the cluster properties in quantum field theories which cannot be 
regarded as deformation of conformal models. A complete understanding of all 
these aspects of the Form Factors would allow us to better understand the 
asymptotic high--energy regime of quantum theories and their operator 
content.

\vspace{3mm}

{\em Acknowledgements.} We are grateful to G. Delfino and P. Simonetti 
for useful discussions. 

\newpage


\appendix
\appsection

In this appendix we give the general parameterization adopted throughout 
the paper for two--particle Form Factors. The $S$--matrices of the 
specific models analysed in this paper are given in Appendix B 
where the generic amplitude 
\be  
\label{Sab}
S_{ab}\,(\th) = \prod_{x \in P_{ab}} (x)^{p_x}  \: 
                \prod_{y \in Z_{ab}} (-y)^{q_y}
\ee
is written adopting the notation 
\be \label{(x)}
(\alpha) = \frac{\tanh\frac{1}{2}\left(\th + i \pi \alpha \right)}
{\tanh\frac{1}{2}\left(\th - i \pi  \alpha\right)}\,\, 
\ee
and the positive rational indices $x$ and $y$ label the poles and
the zeros displayed by the amplitude in the physical strip 
$\mbox{\rm Im\,} \th \in [0, \pi]$.
The bound state simple poles of the $S$--matrices are identified by 
superscripts which denote the particles produced. There are however 
also simple poles of a different nature which have been labelled 
with suffices ${\cal B}$ and ${\cal D}$ that are not related to 
any bound state and are due to multiparticle scattering processes of the 
kind shown in Figure 2 and 3 respectively. The fact that these diagrams 
(which usually produce second and third order poles) are here responsible 
for simple poles is due to the occurrence of zeros in the $S$--matrix 
factors carried by the internal crossing lines \cite{generalized}. 
Higher order poles are present as well and, among these, in 
the model $[{\cal M}(2/9)]_{(1,4)}$, some triple poles labelled by 
an asterisk which also have a non--standard diagrammatic interpretation.
The understanding of the nature of all the poles is necessary in order 
to assign to the FF's the correct pole structure.

The general two--particle Form Factor of a scalar operator $\Phi(x)$ 
\be
   F_{ab}^\Phi\,(\th_1 - \th_2) = \langle \,0\, |\, \Phi (0) \,|
    \, A_a(\th_1)\, A_b(\th_2)\,\rangle  \,\,,
\ee
will be parameterized by 
\be  \label{Fab}
F_{ab}^\Phi\,(\th) = Q_{ab}^\Phi\,(\th) 
\frac{F_{ab}^{\rm min}\,(\th)}{D_{ab}\,(\th)}\,\,.
\ee
where the ``minimal'' FF
\be \label{Fmin}
  F_{ab}^{\rm min}\,(\th) = \Big( -i \sinh \left(\th/2 \right) 
                     \Big)^{\frac{1- S_{ab}(0)}{2}}\:
                      \frac{\bd \prod_{x \in P} g_x^{p_x}(\th)\ed }
                    {\bd \prod_{y \in Z_{ab}} g_y^{q_y}(\th) \ed} \,\,,
\ee
which has neither zeros nor poles in the physical strip, is written in 
terms of the function
\be
g_{x}(\theta)= \prod_{k=0}^{\infty} 
\left[  \frac{  \left[1 + \left[\frac{\frac{i\pi-\th}{2 \pi}}
             {n + \frac{1}{2}+ \frac{x}{2}}  \right]^2 \right] 
                \left[1 + \left[\frac{\frac{i\pi-\th}{2 \pi}}
             {n +     1      - \frac{x}{2}}  \right]^2 \right] }
             {  \left[1 + \left[\frac{\frac{i\pi-\th}{2 \pi}}
             {n +     1      + \frac{x}{2}}  \right]^2 \right]
                \left[1 + \left[\frac{\frac{i\pi-\th}{2 \pi}}
             {n + \frac{3}{2}- \frac{x}{2}}  \right]^2 \right] }
\right]^{k+1}\,.
\ee
This function is normalized by $g_{x}(i\pi) = 1$ and behaves 
asymptotically as
\be 
\label{gas}
g_{x}(\th)\sim e^{|\th|/2}\, 
\hspace{1cm}\mbox{for}\hspace{1cm}\,\th\rightarrow\infty\, .
\ee
The factor $Q_{ab}^\Phi\,(\th)$ in (\ref{Fab}) is a polynomial in 
$\cosh \th$ carrying the dependence on the specific operator $\Phi(x)$ 
\be
\label{qgen}
Q^\Phi_{ab} = \sum_{k=0}^{k_{ab,\Phi}^{max}}  a_{ab,\Phi}^{(k)} 
\cosh^k (\th)   \,\,.
\ee  
The most subtle element in the parameterization of the FF's is represented by 
the structure of the poles which, in eq. (\ref{Fab}) are introduced by 
the factor $D_{ab}(\th)$. In order to establish which poles are to be 
found in a FF one must in general have a complete understanding of 
the nature of the poles in the corresponding $S$--matrix element in 
terms of microscopical processes. We will write in general, 
\be 
\label{Dab}
  D_{ab} \,(\th) = \prod_{x\in P_{ab}} {\cal P}_x^{i_x}\,
                               {\cal P}_{1-x}^{j_x}\,\, ,
\ee
where the set of indices is defined in (\ref{Sab}) 
and 
\be
{\cal P}_x \, (\th) = \frac{\cos (\pi x) - \cosh (\th)}
                           {2 \cos^2 ( \frac{\pi x}{2})} \,\,.
\ee
For bound--state simple poles and ordinary higher order poles of
the $S$--matrix, the correct rule for determining the indices $i_x$ and
$j_x$  is given by \cite{DM} 
\be
\begin{array}{lll}
i_{x} = n\,\,\, , & j_{x} = n-1 \,\,\, , &
\mbox{\rm if} \hspace{1cm} p_x=2n-1\,\,\,; \\
i_{x} = n \,\,\, , & j_{x} = n \,\,\, , &
\mbox{\rm if} \hspace{1cm} p_x=2n\,\,\, ,
\end{array}
\ee
For simple poles of type $(x)_{\cal B}$ and $(x)_{\cal D}$ 
one can show that the correct indices are still 
$i_x=1$ and $j_x=0$, as for a bound state simple pole. 

Notice however that the poles of the FF's induced by the triple poles 
labelled with $*$ in $[{\cal M}(2/9)]_{(1,4)}$ do not fall within 
the above analysis. Their general expressions is not further 
investigated here since these FF's were not needed in the present work. 

As a final remark, notice that every  function $(\alpha)$ could be 
equivalently written as $(1-\alpha)$ without changing the $S$--matrices. 
However, the pole prescription given above for the FF's is {\em sensitive} 
to this change in  the case of odd order poles.
Therefore, all the labels $\alpha$ in the $S$--matrices reported here 
have been chosen to give (in units of $i \pi$) the value of the 
direct $s$--channel resonant angles in the case of bound state 
odd poles and also in the case of poles of type ${\cal B}$ and ${\cal D}$ 
\footnote{For  $s$--channel in these cases we mean the one defined 
by Figures 2 and 3 with particles flowing upward}.
Only with this choice, the above prescription gives
the correct poles of the FF's.

With the parameterization (\ref{Fab}), the two--particle Form Factor 
of a general operator $\Phi$ is therefore completely determined after 
fixing the coefficients $a_{ab,\Phi}^{(k)}$ in the expansion (\ref{qgen}).

\newpage

\appsection
 
In this Appendix we give the $S$--matrices of the integrable models 
analyzed in this work. The function $(\alpha)$ used in the tables is 
given in eq. (\ref{(x)}). Anomalous simple poles have been labelled with 
${\cal B}$ and ${\cal D}$, while the anomalous triple poles of the 
model  $[{\cal M}(2/9)]_{(1,4)}$ are identified with $*$.

\vspace{1cm}

\begin{center}
\begin{tabular}{|c|} \hline
$ 
\begin{array}{ccl}
\rule[-4mm]{0mm}{12mm}S_{11}(\th) & = & \st{\bf 1}{\left(\frac{2}{3}\right)} \,
\st{\bf 2}{\left(\frac{1}{9}\right)} \,\left(-\frac{2}{9}\right)  \\
\rule[-4mm]{0mm}{12mm}S_{12}(\th) & = & \st{\bf 1}{\left(\frac{17}{18}\right)}\,
\left(\frac{11}{18}\right)_{\cal B} \\
\rule[-4mm]{0mm}{12mm}S_{22}(\th) & = & \st{\bf 2}{\left(\frac{2}{3}\right)}
\,\left(\frac{8}{9}\right)_{\cal B} \, \left(\frac{5}{9}\right)_{\cal D} 
\end{array}
$ 
\\ \hline
$
\rule[-5mm]{0mm}{12mm}\bd m_2 = 2\,\cos\frac{\pi}{18}\: m_1 = 1.9696...\: m_1\ed
$ \\
\hline
\end{tabular}
\end{center}
\begin{center}
{\bf Table B1:} S--Matrix and  mass ratios of the $[M(2/7)]_{(1,2)}$ model. 
\end{center}

\vspace{1cm}

\begin{center}
\begin{tabular}{|c|} \hline
$ 
\begin{array}{ccl}
\rule[-4mm]{0mm}{12mm}S_{11}(\th) & = & \st{\bf 2}{\left(\frac{2}{5}\right)} \\
\rule[-4mm]{0mm}{12mm}S_{12}(\th) & = & \st{\bf 1}{\left(\frac{4}{5}\right)}\,\st{\bf 2}{\left(\frac{3}{5}\right)} \\
\rule[-4mm]{0mm}{12mm}S_{22}(\th) & = & \st{\bf 1}{\left(\frac{4}{5}\right)}\,\left(\frac{2}{5}\right)^2 
\end{array}
$ 
\\ \hline
$
\rule[-5mm]{0mm}{12mm}\bd m_2 = 2\,\cos\frac{\pi}{5}\: m_1 = 1.61803 \ldots  m_1\ed
$ \\
\hline
\end{tabular}
\end{center}
\begin{center}
{\bf Table B2:} S--Matrix and  mass ratios of the $[M(2/7)]_{(1,3)}$ model. 
\end{center}

\vspace{1cm}

\begin{center}
\begin{tabular}{|c|} \hline
$ 
\begin{array}{ccl}
\rule[-4mm]{0mm}{12mm}S_{11}(\th) & = & \st{\bf 1}{\left(\frac{2}{3}\right)} \,
\st{\bf 2}{\left(\frac{1}{12}\right)} \,\left(-\frac{1}{4}\right)  \\
\rule[-4mm]{0mm}{12mm}S_{12}(\th) & = & \st{\bf 1}{\left(\frac{23}{24}\right)}\,
\st{\bf 3}{\left(\frac{1}{8}\right)} \,
\left(\frac{5}{8}\right)_{\cal B} \,\left(-\frac{5}{24}\right)  \\
\rule[-4mm]{0mm}{12mm}S_{13}(\th) & = &\st{\bf 2}{\left(\frac{11}{12}\right)}
\,\left(\frac{7}{12}\right)_{\cal B}\\
\rule[-4mm]{0mm}{12mm}S_{22}(\th) & = & \st{\bf 2}{\left(\frac{2}{3}\right)} \, 
\left(\frac{11}{12}\right)^2 \, \left(\frac{7}{12}\right)_{\cal D} 
\, \left(-\frac{1}{4}\right) \\
\rule[-4mm]{0mm}{12mm}S_{23}(\th) & = &\st{\bf 1}{\left(\frac{23}{24}\right)}
\,\left(\frac{7}{8}\right)_{\cal B}
\,\left(\frac{5}{8}\right)_{\cal B}\,\left(\frac{13}{24}\right)_{\cal D} \\
\rule[-4mm]{0mm}{12mm}S_{33}(\th) & = &\st{\bf 3}{\left(\frac{2}{3}\right)}
\,\left(\frac{11}{12}\right)_{\cal B}\,\left(\frac{5}{6}\right)_{\cal B}
\,\left(\frac{7}{12}\right)_{\cal D}\,\left(\frac{1}{2}\right)_{\cal D} \\
\end{array}
$ 
\\ \hline
$
\ba{rcl}
\rule[-6mm]{0mm}{14mm} m_a & = & \bd  \frac{\sin\frac{a \pi}{24}}
{\sin\frac{\pi}{24}}\ed\:  m_1 \:\:\:\:\:\:  a= 1,\, 2,\, 3\\
\ea
$ \\
\hline
\end{tabular}
\end{center}
\begin{center}
{\bf Table B3:} S--Matrix and  mass ratios of the $[M(2/9)]_{(1,2)}$ model. 
\end{center}

\vspace{1cm}

\begin{center}
\begin{tabular}{|c|} \hline
$ 
\begin{array}{ccl}
\rule[-4mm]{0mm}{12mm}S_{11}(\th) & = & \st{\bf 2}{\left(\frac{2}{7}\right)} \\
\rule[-4mm]{0mm}{12mm}S_{12}(\th) & = & \st{\bf 1}{\left(\frac{6}{7}\right)} \, 
                                        \st{\bf 3}{\left(\frac{3}{7}\right)} \\
\rule[-4mm]{0mm}{12mm}S_{13}(\th) & = & \st{\bf 2}{\left(\frac{5}{7}\right)} \, 
                                        \st{\bf 3}{\left(\frac{4}{7}\right)} \\
\rule[-4mm]{0mm}{12mm}S_{22}(\th) & = & \st{\bf 3}{\left(\frac{4}{7}\right)} \, 
                                        \left(\frac{5}{7}\right)^2 \\
\rule[-4mm]{0mm}{12mm}S_{23}(\th) & = & \st{\bf 1}{\left(\frac{6}{7}\right)} \, 
                                        \st{\bf 2}{\left(\frac{5}{7}\right)} \, 
                                        \left(\frac{3}{7}\right)^2 \\
\rule[-4mm]{0mm}{12mm}S_{33}(\th) & = & \st{\bf 1}{\left(\frac{6}{7}\right)} \, 
                                        \left(\frac{3}{7}\right)^2 \,
                                        \left(\frac{5}{7}\right)^2    \\
\end{array}
$ 
\\ \hline
$
\ba{rcl}
\rule[-6mm]{0mm}{14mm} m_a & = & \bd  \frac{\sin\frac{a \pi}{7}}
{\sin\frac{\pi}{7}}\ed\:  m_1 \:\:\:\:\:\:  a= 1,\, 2,\, 3\\
\ea
$ \\
\hline
\end{tabular}
\end{center}
\begin{center}
{\bf Table B4:} S--Matrix and  mass ratios of the $[M(2/9)]_{(1,3)}$ model. 
\end{center}

\vspace{1cm}

\begin{center}
\begin{tabular}{|c|} \hline
$ 
\begin{array}{ccl}
\rule[-4mm]{0mm}{12mm}S_{11}(\th) & = & \st{\bf 1}{\left(\frac{2}{3}\right)} \,
                                        \st{\bf 2}{\left(\frac{7}{15}\right)} \, 
                                        \st{\bf 3}{\left(\frac{2}{15}\right)} \,
                                        \left(-\frac{1}{15}\right) \,
                                        \left(-\frac{2}{5}\right) \\
\rule[-4mm]{0mm}{12mm}S_{12}(\th) & = & \st{\bf 1}{\left(\frac{23}{30}\right)} \,
                                        \st{\bf 3}{\left(\frac{13}{30}\right)}  \\
\rule[-4mm]{0mm}{12mm}S_{13}(\th) & = & \st{\bf 1}{\left(\frac{14}{15}\right)} \,
                                        \st{\bf 2}{\left(\frac{11}{15}\right)} \,
                                        \st{\bf 4}{\left(\frac{1}{5}\right)}   \,
                                        \left(\frac{3}{5}\right)^2   \,
                                        \left(-\frac{2}{15}\right)   \,
                                        \left(-\frac{1}{3}\right)  \\
\rule[-4mm]{0mm}{12mm}S_{14}(\th) & = & \st{\bf 3}{\left(\frac{13}{15}\right)}   \,
                                        \left(\frac{8}{15}\right)_{\cal B}   \,
                                        \left(\frac{2}{3}\right)^2  \\
\rule[-4mm]{0mm}{12mm}S_{22}(\th) & = & \st{\bf 2}{\left(\frac{2}{3}\right)}   \,
                                        \st{\bf 4}{\left(\frac{1}{5}\right)}   \,
                                        \left(\frac{8}{15}\right)_{\cal B}   \\
\rule[-4mm]{0mm}{12mm}S_{23}(\th) & = & \st{\bf 1}{\left(\frac{5}{6}\right)}   \,
                                        \left(\frac{1}{2}\right)_{\cal B}  \,
                                        \left(\frac{7}{10}\right)_{\cal B} \,
                                        \left(\frac{11}{30}\right)_{\cal B}  \\
\rule[-4mm]{0mm}{12mm}S_{24}(\th) & = & \st{\bf 2}{\left(\frac{9}{10}\right)}   \,
                                        \left(\frac{23}{30}\right)_{\cal B}  \,
                                        \left(\frac{3}{10}\right)_{\cal B} \,
                                        \left(\frac{19}{30}\right)_{\cal B}\,
                                        \left(\frac{17}{30}\right)^2  \\
\rule[-4mm]{0mm}{12mm}S_{33}(\th) & = & \st{\bf \!\!\! 3}{\left(\frac{2}{3}\right)^3}   \,
                                        \left(\frac{2}{15}\right)^2  \,
                                        \left(\frac{7}{15}\right)^2  \,
                                        \left(-\frac{1}{15}\right)   \,
                                        \left(-\frac{2}{5}\right)    \\
\rule[-4mm]{0mm}{12mm}S_{34}(\th) & = & \st{\bf  1}{\left(\frac{14}{15}\right)}   \,
                                        \left(\frac{4}{5}\right)_{\cal B}  \,
                                        \left(\frac{7}{15}\right)_{\cal D}  \,
                                        \left(\frac{11}{15}\right)^2   \,
                                        \left(\frac{3}{5}\right)^3_*    \\
\rule[-4mm]{0mm}{12mm}S_{44}(\th) & = & \st{\bf \!\!\! 4}{\left(\frac{2}{3}\right)^3_*}   \,
                                        \left(\frac{8}{15}\right)^3_* \,
                                        \left(\frac{2}{5}\right)_{\cal D}   \,
                                        \left(\frac{11}{15}\right)_{\cal B}  \,
                                        \left(\frac{13}{15}\right)_{\cal B}   \,
                                        \left(\frac{1}{5}\right)^2    
\end{array}
$ 
\\ \hline               
$                     
\ba{rcl}
\rule[-6mm]{0mm}{14mm} m_2 & = & \bd 2 \cos\frac{7 \pi}{30}\: m_1 = 1.48629...\: m_1 \ed \\
\rule[-6mm]{0mm}{14mm} m_3 & = & \bd 2 \cos\frac{ \pi}{15}\: m_1  = 1.95630...\: m_1 \ed \\
\rule[-6mm]{0mm}{14mm} m_4 & = & \bd 2 \cos\frac{\pi}{10}\: m_2   = 2.82709...\: m_1 \ed
\ea
$ \\
\hline
\end{tabular}
\end{center}
\begin{center}
{\bf Table B5:} S--Matrix and  mass ratios of the $[M(2/9)]_{(1,4)}$ model. 
\end{center}


\newpage

\vspace{25mm}

{\bf Table Captions}

\vspace{1cm}

\begin{description} 
\item [Table 1] One--particle form factors of cluster solutions in  $[{\cal M}(2/7)]_{(1,2)}$.
\item [Table 2] Two--particle form factors coefficients of cluster solutions in $[{\cal M}(2/7)]_{(1,2)}$.
\item [Table 3] One--particle form factors of cluster solutions in  $[{\cal M}(2/7)]_{(1,3)}$.
\item [Table 4] Two--particle form factors coefficients of cluster solutions in $[{\cal M}(2/7)]_{(1,3)}$.
\item [Table 5] Sum rules of the conformal dimensions of primary operators in $[{\cal M}(2/7)]_{(1,2)}$. 
\item [Table 6] Sum rules of the conformal dimensions of primary operators in $[{\cal M}(2/7)]_{(1,2)}$. 
\item [Table 7] One--particle form factors of cluster solutions in  $[{\cal M}(2/9)]_{(1,2)}$.
\item [Table 8] Two--particle form factors coefficients of cluster solutions in $[{\cal M}(2/9)]_{(1,2)}$.
\item [Table 9] One--particle form factors of cluster solutions in  $[{\cal M}(2/9)]_{(1,3)}$.
\item [Table 10] Two--particle form factors coefficients of cluster solutions in $[{\cal M}(2/9)]_{(1,3)}$.
\item [Table 11] One--particle form factors of cluster solutions in  $[{\cal M}(2/9)]_{(1,4)}$.
\item [Table 12] Two--particle form factors coefficients of cluster solutions in $[{\cal M}(2/9)]_{(1,4)}$.
\item [Table 13] Sum rules of the conformal dimensions of primary operators in $[{\cal M}(2/9)]_{(1,2)}$.
\item [Table 14] Sum rules of the conformal dimensions of primary operators in $[{\cal M}(2/9)]_{(1,3)}$.
\item [Table 15] Sum rules of the conformal dimensions of primary operators in $[{\cal M}(2/9)]_{(1,4)}$.
\item [Table 16] Comparison between numerical and theoretical estimates of data obtained in different 
                  non--integrable deformations of ${\cal M}_{(2,9)}$.
                 .

\end{description}

\newpage

\begin{center}
\begin{tabular}{||c|r|r||} \hline
\rule[-2mm]{0mm}{7mm}${\cal O}$& \multicolumn{1}{c|}{$\phi_{1,2} $ }  
                               & \multicolumn{1}{c||}{$\phi_{1,3}$}\\ \hline \hline
\rule[-2mm]{0mm}{7mm}$F_1^{\cal O}$ & $0.8129447456 \:i $   & $1.245503611 \:i$\\ \hline
\rule[-2mm]{0mm}{7mm}$F_2^{\cal O}$ & $-0.1200387686  $   &  $-0.4656766285$\\ \hline
\end{tabular}
\end{center}
\begin{center}
{\bf Table 1}
\end{center}
\vspace{5mm}
\begin{center}
\begin{tabular}{||c|r|r||} \hline
\rule[-2mm]{0mm}{7mm}${\cal O}$& \multicolumn{1}{c|}{$\phi_{1,2} $ }  
                               & \multicolumn{1}{c||}{$\phi_{1,3}$}\\ \hline \hline
\rule[-2mm]{0mm}{7mm}$a_{11,{\cal O}}^{(0)}$ & $ -0.6905355776$  & $-0.4178217785$\\ 
\rule[-2mm]{0mm}{7mm}$a_{11,{\cal O}}^{(1)}$ & $  1.570496171$   & $3.686419944$\\ \hline
\rule[-2mm]{0mm}{7mm}$a_{12,{\cal O}}^{(0)}$ & $ 31.91217166 \:i$& $160.8200658 \:i$\\ 
\rule[-2mm]{0mm}{7mm}$a_{12,{\cal O}}^{(1)}$ & $ 25.76337182\: i$& $153.1262244 \:i$\\ \hline
\rule[-2mm]{0mm}{7mm}$a_{22,{\cal O}}^{(0)}$ & $-12.74804909$    & $-71.66459155$\\ 
\rule[-2mm]{0mm}{7mm}$a_{22,{\cal O}}^{(1)}$ & $ -3.97663589$    & $-59.84689851$\\ \hline
\end{tabular}
\end{center}
\begin{center}
{\bf Table 2}
\end{center}
\vspace{5mm}
\begin{center}
\begin{tabular}{||c|r|r||} \hline
\rule[-2mm]{0mm}{7mm}${\cal O}$& \multicolumn{1}{c|}{$\phi_{1,2} $ }  
                               & \multicolumn{1}{c||}{$\phi_{1,3}$}\\ \hline \hline
\rule[-2mm]{0mm}{7mm}$F_1^{\cal O}$ & $ 0.8703387193\:i $   & $  1.408237641 \:i$\\ \hline
\rule[-2mm]{0mm}{7mm}$F_2^{\cal O}$ & $-0.3322661173  $   &   $ -0.8698840033$\\ \hline
\end{tabular}
\end{center}
\begin{center}
{\bf Table 3}
\end{center}
\vspace{5mm}
\begin{center}
\begin{tabular}{||c|r|r||} \hline
\rule[-2mm]{0mm}{7mm}${\cal O}$& \multicolumn{1}{c|}{$\phi_{1,2} $ }  
                               & \multicolumn{1}{c||}{$\phi_{1,3}$}\\ \hline \hline
\rule[-2mm]{0mm}{7mm}$a_{11,{\cal O}}^{(0)}$ & $-1.453085043$     & $-3.804226098$\\ \hline
\rule[-2mm]{0mm}{7mm}$a_{12,{\cal O}}^{(0)}$ & $ 10.38924846\:i$  & $ 30.40986050\:i$\\ 
\rule[-2mm]{0mm}{7mm}$a_{12,{\cal O}}^{(1)}$ & $ 6.420908640\:i$  & $ 27.19940617\:i$\\ \hline
\rule[-2mm]{0mm}{7mm}$a_{22,{\cal O}}^{(0)}$ & $-13.76381909$     & $-42.18951412$\\ 
\rule[-2mm]{0mm}{7mm}$a_{22,{\cal O}}^{(1)}$ & $-4.702281947$     & $-32.22992104$\\ \hline
\end{tabular}
\end{center}
\begin{center}
{\bf Table 4}
\end{center}
\begin{center}
\begin{tabular}{||c|r|r|r||} \hline
\rule[-2mm]{0mm}{7mm} {\it states}& \multicolumn{1}{c|}{\it s}
                                  & \multicolumn{1}{c|}{\it $\Delta_{12}$--terms}  
                                  & \multicolumn{1}{c||}{\it $\Delta_{13}$--terms}\\ \hline \hline
\rule[-2mm]{0mm}{7mm}$A_1$             & $1.000\,m_1$        & $-0.2922910 $  & $-0.4478157$\\ 
\rule[-2mm]{0mm}{7mm}$A_2$             & $1.969\,m_1$        & $ 0.0016428 $  & $ 0.0063729$\\ 
\rule[-2mm]{0mm}{7mm}$A_1$ $A_1$       & $\geq 2.000\,m_1$   & $ 0.0051123 $  & $ 0.0137590$\\ 
\rule[-2mm]{0mm}{7mm}$A_1$ $A_2$       & $\geq 2.969\,m_1$   & $-0.0000763 $  & $-0.0004400$\\ 
\rule[-2mm]{0mm}{7mm}$A_1$ $A_1$ $A_1$ & $\geq 3.000\,m_1$   & $-0.0001040 $  & $-0.0004777$\\ 
\rule[-2mm]{0mm}{7mm}$A_2$ $A_2$       & $\geq 3.939\,m_1$   & $ 0.0000003 $  & $ 0.0000040$\\ \hline\hline
\multicolumn{2}{||c|}{\rule[-2mm]{0mm}{7mm} \it sum }       & $-0.2857159 $  & $-0.4285976$\\ \hline
\multicolumn{2}{||c|}{\rule[-2mm]{0mm}{7mm} \it value expected}& $ -0.2857143 $  & $ -0.4285714$\\ \hline\hline
\end{tabular}
\end{center}
\begin{center}
{\bf Table 5}
\end{center}
\vspace{5mm}
\begin{center}
\begin{tabular}{||c|r|r|r||} \hline
\rule[-2mm]{0mm}{7mm} {\it states}& \multicolumn{1}{c|}{\it s}
                                  & \multicolumn{1}{c|}{\it $\Delta_{12}$--terms}  
                                  & \multicolumn{1}{c||}{\it $\Delta_{13}$--terms}\\ \hline \hline
\rule[-2mm]{0mm}{7mm}$A_1$             & $1.000\,m_1$ & $-0.3221795 $  & $-0.5212974$\\ 
\rule[-2mm]{0mm}{7mm}$A_2$             & $1.618\,m_1$ & $ 0.0290206 $  & $ 0.0759768$\\ 
\rule[-2mm]{0mm}{7mm}$A_1$ $A_1$       & $2.000\,m_1$ & $ 0.0098699 $  & $ 0.0258398$\\ 
\rule[-2mm]{0mm}{7mm}$A_1$ $A_2$       & $2.618\,m_1$ & $-0.0023149 $  & $-0.0089996$\\ 
\rule[-2mm]{0mm}{7mm}$A_1$ $A_1$ $A_1$ & $3.000\,m_1$ & $-0.0003334 $  & $-0.0013803$\\
\rule[-2mm]{0mm}{7mm}$A_2$ $A_2$       & $3.236\,m_1$ & $ 0.0001155 $  & $ 0.0006612$\\ \hline
\multicolumn{2}{||c|}{\rule[-2mm]{0mm}{7mm} \it sum } & $-0.2858218 $  & $-0.4291998$\\ \hline
\multicolumn{2}{||c|}{\rule[-2mm]{0mm}{7mm} \it value expected}& $-0.2857143  $  & $ -0.4285714$\\ \hline\hline
\end{tabular}
\end{center}
\begin{center}
{\bf Table 6}
\end{center}
\vspace{5mm}
\begin{center}
\begin{tabular}{||c|r|r|r||} \hline
\rule[-2mm]{0mm}{7mm}${\cal O}$& \multicolumn{1}{c|}{$\phi_{1,2}$}  
                               & \multicolumn{1}{c|}{$\phi_{1,3}$}
                               & \multicolumn{1}{c||}{$\phi_{1,4}$} \\ \hline \hline
\rule[-2mm]{0mm}{7mm}$F_1^{\cal O}$ & $ 0.7548301717 \:i$   & $1.288575652 \:i$ & $1.564862744\:i$ \\ \hline
\rule[-2mm]{0mm}{7mm}$F_2^{\cal O}$ & $-0.1056909725$   & $-0.4593398099$ & $-0.7331609072$ \\ \hline
\rule[-2mm]{0mm}{7mm}$F_3^{\cal O}$ & $-0.01375684037 \:i$   & $-0.1175389994 \:i$ & $-0.2854817817\:i$ \\ \hline
\end{tabular}
\end{center}
\begin{center}
{\bf Table 7}
\end{center}
\begin{center}
\begin{tabular}{||c|r|r|r||} \hline
\rule[-2mm]{0mm}{7mm}${\cal O}$& \multicolumn{1}{c|}{$\phi_{1,2}$}  
                               & \multicolumn{1}{c|}{$\phi_{1,3}$}
                               & \multicolumn{1}{c||}{$\phi_{1,4}$} \\ \hline \hline
\rule[-2mm]{0mm}{7mm}$a_{11,{\cal O}}^{(0)}$ & $-0.3810248990$   & $0.1280888115$    & $0.6449629545$ \\
\rule[-2mm]{0mm}{7mm}$a_{11,{\cal O}}^{(1)}$ & $1.289925788$     & $3.759118917$     & $5.543942595$ \\ \hline
\rule[-2mm]{0mm}{7mm}$a_{12,{\cal O}}^{(0)}$ & $14.10905183 \:i$ & $ 75.18632019\:i$ & $110.3472056 \:i$ \\
\rule[-2mm]{0mm}{7mm}$a_{12,{\cal O}}^{(1)}$ & $-12.74323779\:i$ & $-79.90895489\:i$ & $-180.9845092 \:i$ \\
\rule[-2mm]{0mm}{7mm}$a_{12,{\cal O}}^{(2)}$ & $-19.36998044\:i$ & $-143.7096872\:i$ & $-278.5592522 \:i$ \\ \hline
\rule[-2mm]{0mm}{7mm}$a_{13,{\cal O}}^{(0)}$ & $-1.826322080$    & $-18.97540047$    & $-51.56786333$ \\
\rule[-2mm]{0mm}{7mm}$a_{13,{\cal O}}^{(1)}$ & $-1.116015559$    & $-16.27774386$    & $-48.01279071$ \\ \hline
\rule[-2mm]{0mm}{7mm}$a_{22,{\cal O}}^{(0)}$ & $-1.466545085$    & $-3.003367424$    & $14.9160654$ \\
\rule[-2mm]{0mm}{7mm}$a_{22,{\cal O}}^{(1)}$ & $ 7.821352950$    & $ 60.49540624$    & $160.4007705$ \\
\rule[-2mm]{0mm}{7mm}$a_{22,{\cal O}}^{(2)}$ & $ 2.717967823$    & $ 51.33773403$    & $130.7877664$ \\ \hline
\rule[-2mm]{0mm}{7mm}$a_{23,{\cal O}}^{(0)}$ & $ 153.8279467 \:i$& $1842.946063\:i$  & $5426.66381 \:i$ \\
\rule[-2mm]{0mm}{7mm}$a_{23,{\cal O}}^{(1)}$ & $ 175.5584268 \:i$& $2962.508857\:i$  & $9796.436391 \:i$ \\
\rule[-2mm]{0mm}{7mm}$a_{23,{\cal O}}^{(2)}$ & $ 30.43124786 \:i$& $1130.002086\:i$  & $4380.673323 \:i$ \\ \hline
\rule[-2mm]{0mm}{7mm}$a_{33,{\cal O}}^{(0)}$ & $-32.42110324$    & $-450.0936155$    & $-1394.808207$ \\
\rule[-2mm]{0mm}{7mm}$a_{33,{\cal O}}^{(1)}$ & $-20.23293766$    & $-589.1376530$    & $-2309.626757$ \\
\rule[-2mm]{0mm}{7mm}$a_{33,{\cal O}}^{(2)}$ & $-2.174915595$    & $-158.7701993$    & $-936.6165096$ \\ \hline
\end{tabular}
\end{center}
\begin{center}
{\bf Table 8}
\end{center}

\vspace{1cm}

\begin{center}
\begin{tabular}{||c|r|r|r||} \hline
\rule[-2mm]{0mm}{7mm}${\cal O}$& \multicolumn{1}{c|}{$\phi_{1,2}$}  
                               & \multicolumn{1}{c|}{$\phi_{1,3}$}
                               & \multicolumn{1}{c||}{$\phi_{1,4}$} \\ \hline \hline
\rule[-2mm]{0mm}{7mm}$F_1^{\cal O}$ & $0.8020765716 \:i$    & $1.445292066 \:i$   & $1.802249672\:i$ \\ \hline
\rule[-2mm]{0mm}{7mm}$F_2^{\cal O}$ & $-0.3139111339  $     & $ -1.019263084 $    & $-1.584911324$ \\ \hline
\rule[-2mm]{0mm}{7mm}$F_3^{\cal O}$ & $-0.1373692453 \:i$   & $-0.5561967434 \:i$ & $-1.002231818\:i$ \\ \hline
\end{tabular}
\end{center}
\begin{center}
{\bf Table 9}
\end{center}

\vspace{1cm}

\begin{center}
\begin{tabular}{||c|r|r|r||} \hline
\rule[-2mm]{0mm}{7mm}${\cal O}$& \multicolumn{1}{c|}{$\phi_{1,2}$}  
                               & \multicolumn{1}{c|}{$\phi_{1,3}$}
                               & \multicolumn{1}{c||}{$\phi_{1,4}$} \\ \hline \hline
\rule[-2mm]{0mm}{7mm}$a_{11,{\cal O}}^{(0)}$ & $-0.9631492344$     & $-3.127326026$      & $-4.862860736$ \\\hline
\rule[-2mm]{0mm}{7mm}$a_{12,{\cal O}}^{(0)}$ & $  10.64696613 \:i$   & $ 40.73951464 \:i$    & $72.35568181 \:i$ \\
\rule[-2mm]{0mm}{7mm}$a_{12,{\cal O}}^{(1)}$ & $  5.908620424 \:i$   & $ 34.57048356 \:i$    & $67.03219861 \:i$ \\\hline
\rule[-2mm]{0mm}{7mm}$a_{13,{\cal O}}^{(0)}$ & $ -2.592348236$     & $-11.32977918$      & $-21.24912975$ \\
\rule[-2mm]{0mm}{7mm}$a_{13,{\cal O}}^{(1)}$ & $ -1.153703500$     & $-8.417302355$      & $-18.91350458$ \\\hline
\rule[-2mm]{0mm}{7mm}$a_{22,{\cal O}}^{(0)}$ & $ -5.978990567$     & $-26.44069921$      & $-49.87674876$ \\
\rule[-2mm]{0mm}{7mm}$a_{22,{\cal O}}^{(1)}$ & $ -1.544771430$     & $-16.28633559$      & $-39.37864116$ \\\hline
\end{tabular}
\end{center}
\begin{center}
{\bf Table 10}
\end{center}

\vspace{1cm}
\begin{center}
\begin{tabular}{||c|r|r|r||} \hline
\rule[-2mm]{0mm}{7mm}${\cal O}$& \multicolumn{1}{c|}{$\phi_{1,2}$}  
                               & \multicolumn{1}{c|}{$\phi_{1,3}$}
                               & \multicolumn{1}{c||}{$\phi_{1,4}$} \\ \hline \hline
\rule[-2mm]{0mm}{7mm}$F_1^{\cal O}$ & $-0.9043544898 \: i$ & $-1.72785339 \: i$  & $-2.211259663 \: i$ \\ \hline
\rule[-2mm]{0mm}{7mm}$F_2^{\cal O}$ & $-0.5483648961$      & $-1.476188315$      & $-2.169493373$      \\ \hline
\rule[-2mm]{0mm}{7mm}$F_3^{\cal O}$ & $0.2673316508 \: i$  & $0.8709319528 \: i$ & $1.45902371 \: i$   \\ \hline
\rule[-2mm]{0mm}{7mm}$F_4^{\cal O}$ & $-0.08488118964$     & $-0.3489749771$     & $-0.6451795597$     \\ \hline
\end{tabular}
\end{center}
\begin{center}
{\bf Table 11}
\end{center}

\vspace{1cm}

\begin{center}
\begin{tabular}{||c|r|r|r||} \hline
\rule[-2mm]{0mm}{7mm}${\cal O}$& \multicolumn{1}{c|}{$\phi_{1,2}$}  
                               & \multicolumn{1}{c|}{$\phi_{1,3}$}
                               & \multicolumn{1}{c||}{$\phi_{1,4}$} \\ \hline \hline
\rule[-2mm]{0mm}{7mm}$a_{11,{\cal O}}^{(0)}$ & $1.623982681$       & $4.256426530$     & $6.219867507$ \\
\rule[-2mm]{0mm}{7mm}$a_{11,{\cal O}}^{(1)}$ & $-0.9778411563$     & $-1.325966569$   & $-1.477684504$ \\ 
\rule[-2mm]{0mm}{7mm}$a_{11,{\cal O}}^{(2)}$ & $-2.029027259$      & $-7.406691607$   & $-12.13081476$ \\ \hline
\rule[-2mm]{0mm}{7mm}$a_{12,{\cal O}}^{(0)}$ & $-9.935037127 \: i$    & $-30.48935000 \: i$    & $-50.02403946 \: i$ \\
\rule[-2mm]{0mm}{7mm}$a_{12,{\cal O}}^{(1)}$ & $-4.790105254 \: i$    & $-24.63686052 \: i$ & $-46.33773309 \: i$ \\ \hline
\rule[-2mm]{0mm}{7mm}$a_{13,{\cal O}}^{(0)}$ & $-45.21074197$      & $-145.9600730$   & $-220.9609710$ \\
\rule[-2mm]{0mm}{7mm}$a_{13,{\cal O}}^{(1)}$ & $-441.3756086$      & $-1583.189947$   & $-2731.357697$ \\ 
\rule[-2mm]{0mm}{7mm}$a_{13,{\cal O}}^{(2)}$ & $-533.3237140$      & $-2301.408527$   & $-4364.089257$ \\
\rule[-2mm]{0mm}{7mm}$a_{13,{\cal O}}^{(3)}$ & $-139.4173540$      & $-867.7984505$   & $-1860.500753$ \\ \hline
\rule[-2mm]{0mm}{7mm}$a_{14,{\cal O}}^{(0)}$ & $44.34032961 \: i$     & $189.6077348 \: i$  & $357.8357762 \: i$ \\
\rule[-2mm]{0mm}{7mm}$a_{14,{\cal O}}^{(1)}$ & $54.79194008 \: i$     & $275.9507675 \: i$  & $562.9638801 \: i$ \\ 
\rule[-2mm]{0mm}{7mm}$a_{14,{\cal O}}^{(2)}$ & $11.43395882 \: i$     & $89.81474554 \: i$  & $212.5038539 \: i$ \\ \hline
\rule[-2mm]{0mm}{7mm}$a_{22,{\cal O}}^{(0)}$ & $-9.190266093$      & $-30.91094092$   & $-52.20534546$ \\
\rule[-2mm]{0mm}{7mm}$a_{22,{\cal O}}^{(1)}$ & $-2.709639668$      & $-19.63612453$   & $-42.41201502$ \\ \hline
\rule[-2mm]{0mm}{7mm}$a_{23,{\cal O}}^{(0)}$ & $-81.75802420 \: i$    & $-304.2244838 \: i$ & $-530.8872409 \: i$ \\
\rule[-2mm]{0mm}{7mm}$a_{23,{\cal O}}^{(1)}$ & $-92.61128143 \: i$    & $-446.8601169 \: i$ & $-884.9723034 \: i$ \\
\rule[-2mm]{0mm}{7mm}$a_{23,{\cal O}}^{(2)}$ & $-16.66533965 \: i$    & $-146.1571720 \: i$    & $-359.8444599 \: i$ \\ \hline
\end{tabular}
\end{center}
\begin{center}
{\bf Table 12}
\end{center}

\vspace{1cm}

\begin{center}
\begin{tabular}{||c|r|r|r|r||} \hline
\rule[-2mm]{0mm}{7mm} {\it states}& \multicolumn{1}{c|}{\it s}
                                  & \multicolumn{1}{c|}{\it $\Delta_{12}$--terms}  
                                  & \multicolumn{1}{c|}{\it $\Delta_{13}$--terms}
                                  & \multicolumn{1}{c||}{\it $\Delta_{14}$--terms}
                                  \\ \hline \hline
\rule[-2mm]{0mm}{7mm}$A_1$             & $     1.000\,m_1$        & $-0.3409847 $ & $ -0.5820972$ & $-0.7069063 $ \\ 
\rule[-2mm]{0mm}{7mm}$A_2$             & $     1.982\,m_1$        & $ 0.0017003 $ & $  0.0073894$ & $ 0.0117945 $ \\ 
\rule[-2mm]{0mm}{7mm}$A_1$ $A_1$       & $\geq 2.000\,m_1$        & $ 0.0061957 $ & $  0.0207909$ & $ 0.0316698 $ \\ 
\rule[-2mm]{0mm}{7mm}$A_3$             & $     2.931\,m_1$        & $-0.0000132 $ & $ -0.0001126$ & $-0.0002734 $ \\ 
\rule[-2mm]{0mm}{7mm}$A_1$ $A_2$       & $\geq 2.982\,m_1$        & $-0.0000951 $ & $ -0.0007084$ & $-0.0014392 $ \\ 
\rule[-2mm]{0mm}{7mm}$A_1$ $A_1$ $A_1$ & $\geq 3.000\,m_1$        & $-0.0001421 $ & $ -0.0009038$ & $-0.0017386 $ \\ 
\rule[-2mm]{0mm}{7mm}$A_1$ $A_3$       & $\geq 3.931\,m_1$        & $ 0.0000009 $ & $  0.0000117$ & $ 0.0000339 $ \\ 
\rule[-2mm]{0mm}{7mm}$A_2$ $A_2$       & $\geq 3.965\,m_1$        & $ 0.0000004 $ & $  0.0000061$ & $ 0.0000157 $ \\  
\rule[-2mm]{0mm}{7mm}$A_2$ $A_3$       & $\geq 4.914\,m_1$        & $-0.0000000 $ & $ -0.0000002$ & $-0.0000008 $ \\  \hline\hline
\multicolumn{2}{||c|}{\rule[-2mm]{0mm}{7mm} \it sum }             & $-0.3333379 $ & $ -0.5556241$ & $-0.6668445 $ \\ \hline
\multicolumn{2}{||c|}{\rule[-2mm]{0mm}{7mm} \it value expected}&    $-0.3333333 $ & $ -0.5555556$ & $-0.6666667 $ \\ \hline\hline
\end{tabular}
\end{center}
\begin{center}
{\bf Table 13}
\end{center}

\vspace{1cm}

\begin{center}
\begin{tabular}{||c|r|r|r|r||} \hline
\rule[-2mm]{0mm}{7mm} {\it states}& \multicolumn{1}{c|}{\it s}
                                  & \multicolumn{1}{c|}{\it $\Delta_{12}$--terms}  
                                  & \multicolumn{1}{c|}{\it $\Delta_{13}$--terms}
                                  & \multicolumn{1}{c||}{\it $\Delta_{14}$--terms}
                                  \\ \hline \hline
\rule[-2mm]{0mm}{7mm}$A_1$             & $      1.000\,m_1$        & 
$ -0.370679 $ & $-0.667941 $ & $-0.832909$ \\
 
\rule[-2mm]{0mm}{7mm}$A_2$             & $      1.802\,m_1$        & 
$  0.031509 $ & $ 0.102310 $ & $ 0.159088$ \\
 
\rule[-2mm]{0mm}{7mm}$A_1$  $A_1$      & $ \geq 2.000\,m_1$        & 
$  0.013898 $ & $ 0.045127 $ & $ 0.070170$ \\
 
\rule[-2mm]{0mm}{7mm}$A_3$             & $      2.247\,m_1$        & 
$ -0.004839 $ & $-0.019592 $ & $-0.035304$ \\
 
\rule[-2mm]{0mm}{7mm}$A_1$  $A_2$      & $ \geq 2.802\,m_1$        & 
$ -0.003604 $ & $-0.018722 $ & $-0.035573$ \\

\rule[-2mm]{0mm}{7mm}$A_1$  $A_1$ $A_1$  & $ \geq 3.000\,m_1$        & 
$ -0.000628 $ & $-0.003514 $ & $ -0.006763 $ \\ 

\rule[-2mm]{0mm}{7mm}$A_1$  $A_3$      & $ \geq 3.247\,m_1$        & 
$  0.000663 $ & $ 0.004114 $ & $ 0.008844$ \\
 
\rule[-2mm]{0mm}{7mm}$A_2$  $A_2$      & $ \geq 3.604\,m_1$        & 
$  0.000211 $ & $ 0.001684 $ & $ 0.003864$ \\ \hline\hline
\multicolumn{2}{||c|}{\rule[-2mm]{0mm}{7mm} \it sum }              & 
$ -0.333469 $ & $-0.556534 $ & $-0.668583$ \\ \hline
\multicolumn{2}{||c|}{\rule[-2mm]{0mm}{7mm} \it value expected}&     
$ -0.333333 $ & $-0.555556 $ & $-0.666667$ \\ \hline\hline
\end{tabular}
\end{center}
\begin{center}
{\bf Table 14}
\end{center}

\vspace{1cm}

\begin{center}
\begin{tabular}{||c|r|r|r|r||} \hline
\rule[-2mm]{0mm}{7mm} {\it states}& \multicolumn{1}{c|}{\it s}
                                  & \multicolumn{1}{c|}{\it $\Delta_{12}$--terms}  
                                  & \multicolumn{1}{c|}{\it $\Delta_{13}$--terms}
                                  & \multicolumn{1}{c||}{\it $\Delta_{14}$--terms}
                                  \\ \hline \hline
\rule[-2mm]{0mm}{7mm}$A_1$             & $ 1.000\,m_1$            & $-0.451081$ & $ -0.861833$   & $ -1.102950$ \\ 
\rule[-2mm]{0mm}{7mm}$A_2$             & $ 1.486\,m_1$            & $ 0.121478$ & $  0.327017$   & $  0.480603$ \\ 
\rule[-2mm]{0mm}{7mm}$A_3$             & $ 1.956\,m_1$            & $-0.022989$ & $ -0.074895$   & $ -0.125468$ \\ 
\rule[-2mm]{0mm}{7mm}$A_1$ $A_1$       & $\geq 2.000\,m_1$        & $ 0.035896$ & $  0.121577$   & $  0.197637$ \\ 
\rule[-2mm]{0mm}{7mm}$A_1$ $A_2$       & $\geq 2.486\,m_1$        & $-0.023279$ & $ -0.101138$   & $ -0.183618$ \\ 
\rule[-2mm]{0mm}{7mm}$A_4$             & $ 2.827\,m_1$            & $ 0.001546$ & $  0.006354$   & $  0.011748$ \\ 
\rule[-2mm]{0mm}{7mm}$A_1$ $A_3$       & $\geq 2.956\,m_1$        & $ 0.004304$ & $  0.022374$   & $  0.045474$ \\ 
\rule[-2mm]{0mm}{7mm}$A_2$ $A_2$       & $\geq 2.973\,m_1$        & $-0.001535$ & $ -0.009929$   & $ -0.022429$ \\ 
\rule[-2mm]{0mm}{7mm}$A_2$ $A_3$       & $\geq 3.443\,m_1$        & $-0.000330$ & $ -0.002101$   & $ -0.004686$ \\ 
\rule[-2mm]{0mm}{7mm}$A_1$ $A_4$       & $\geq 3.827\,m_1$        & $ 0.003595$ & $  0.020054$   & $  0.040870$ \\   \hline\hline
\multicolumn{2}{||c|}{\rule[-2mm]{0mm}{7mm} \it sum }             & $-0.332396$ & $ -0.552519$   & $ -0.662819$ \\ \hline
\multicolumn{2}{||c|}{\rule[-2mm]{0mm}{7mm} \it value expected}&    $-0.333333$ & $ -0.555556$   & $ -0.666667$ \\ \hline\hline
\end{tabular}
\end{center}
\begin{center}
{\bf Table 15}
\end{center}

\vspace{1cm}

\begin{center}
\begin{tabular}{||c|r|r|r|r||} \hline \hline
$ \rule[-6mm]{0mm}{14mm} $& \multicolumn{2}{c|}{$\bd \frac{\delta m_1}{\delta m_2}   \ed$}
&  \multicolumn{2}{c||}{$\bd \frac{\delta {\cal E}_{vac}}{m_1^{(0)} \delta m_1} \ed$}   \\ \hline
{\it deformation} & \multicolumn{1}{c|}{\em numerical ($\pm 3\%$)} & \multicolumn{1}{c|}{\em theoretical}
 & \multicolumn{1}{c|}{\em numerical ($\pm 3\%$)} & \multicolumn{1}{c||}{\em theoretical} \\ \hline
$[{\cal M}(2,9)]_{(1,2)} + \varepsilon  \phi_{1,3}$ &$0.590 $&$ 0.592049  $&$-0.275 $&$-0.275404 $ \\ \hline
$[{\cal M}(2,9)]_{(1,2)} + \varepsilon  \phi_{1,4}$ &$0.661 $&$0.660963   $&$-0.204 $&$-0.204124 $ \\ \hline
$[{\cal M}(2,9)]_{(1,3)} + \varepsilon  \phi_{1,2}$ &$0.390 $&$0.391396   $&$-1.04 $&$-1.03826 $ \\ \hline
$[{\cal M}(2,9)]_{(1,3)} + \varepsilon  \phi_{1,4}$ &$0.811 $&$0.83681    $&$-0.205 $&$-0.205640 $ \\ \hline
$[{\cal M}(2,9)]_{(1,4)} + \varepsilon  \phi_{1,2}$ &$-0.133 $&$-0.131367 $&$ 1.73 $&$1.74582 $ \\ \hline
$[{\cal M}(2,9)]_{(1,4)} + \varepsilon  \phi_{1,3}$ &$0.238 $&$0.240486   $&$-0.550 $&$-0.548156 $ \\ \hline\hline
\end{tabular}
\end{center}
\begin{center}
{\bf Table 16}
\end{center}
\vspace{1cm}

\newpage

\vspace{25mm}

{\bf Figure Captions}

\vspace{1cm}

\begin{description}
\item [Figure 1] Bound--state simple pole diagram. 
\item [Figure 2] ``Butterfly'' diagram.
\item [Figure 3] ``Dragonfly'' diagram.
\item [Figure 4] Numerical TCS estimates of $\delta m_1$ versus $\delta m_2$ for
different values of the ``non--integrable'' coupling in the model $[{\cal M}(2/7)]_{(1,2)} + \varepsilon
\phi_{1,3}$. The dashed line represents the theoretical prediction.
\item [Figure 5] Numerical TCS estimates of 
 $\delta {\cal E}_{vac}$ versus $m_1^{(0)} \delta m_1$ for
different values of the ``non--integrable'' coupling in the model $[{\cal M}(2/7)]_{(1,2)} + \varepsilon
\phi_{1,3}$. The dashed line represents the theoretical prediction.
\end{description}

\newpage

\begin{figure}[h]
\vspace{2cm}
\centerline{\hspace{2cm}\psfig{figure=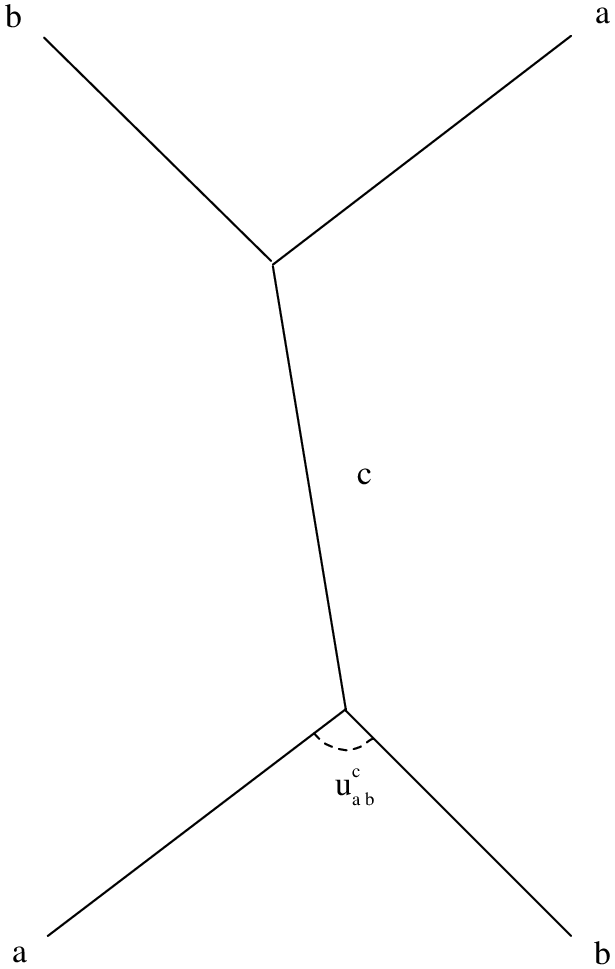,height=5.5in}}
\begin{center}
{\bf Figure 1}
\end{center}
\end{figure}

\begin{figure}[h]
\centerline{\hspace{2cm}\psfig{figure=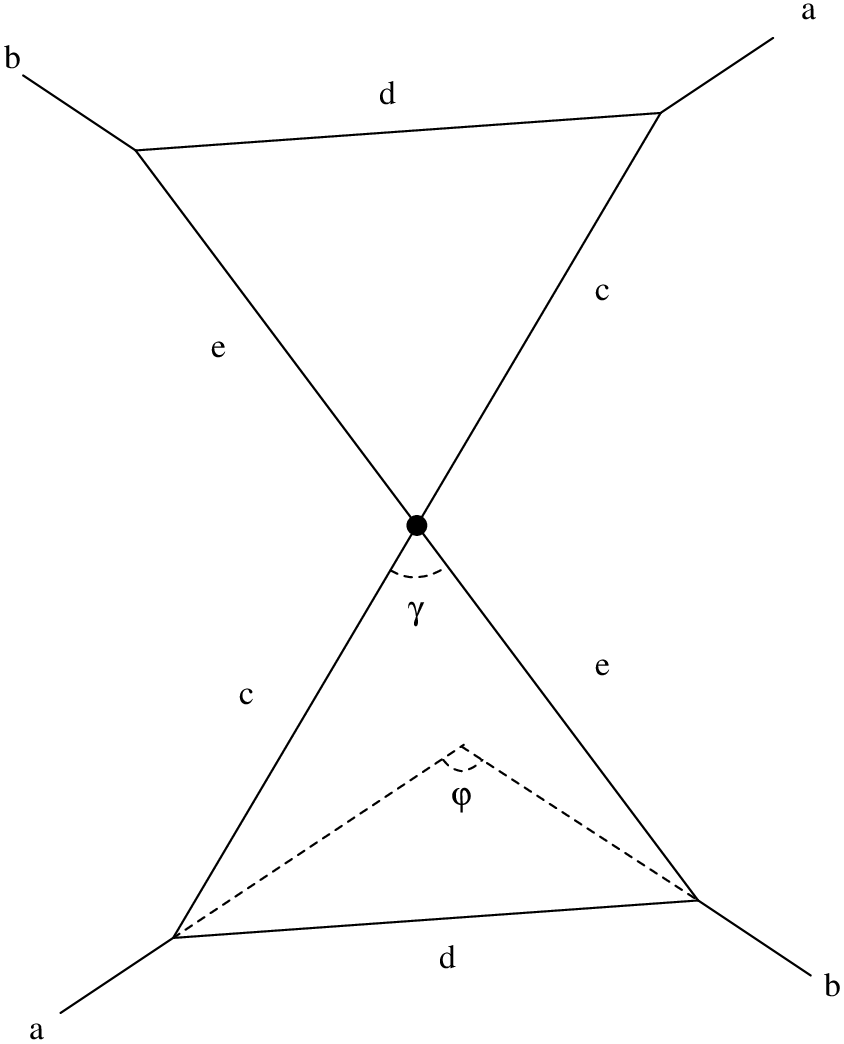,height= 5.5in}}
\begin{center}
{\bf Figure 2}
\end{center}
\end{figure}

\begin{figure}[h]
\centerline{\hspace{2cm}\psfig{figure=dragonfly2.ps,height=6in}}
\begin{center}
{\bf Figure 3}
\end{center}
\end{figure}

\begin{figure}[h]
\centerline{\hspace{2cm}\psfig{figure=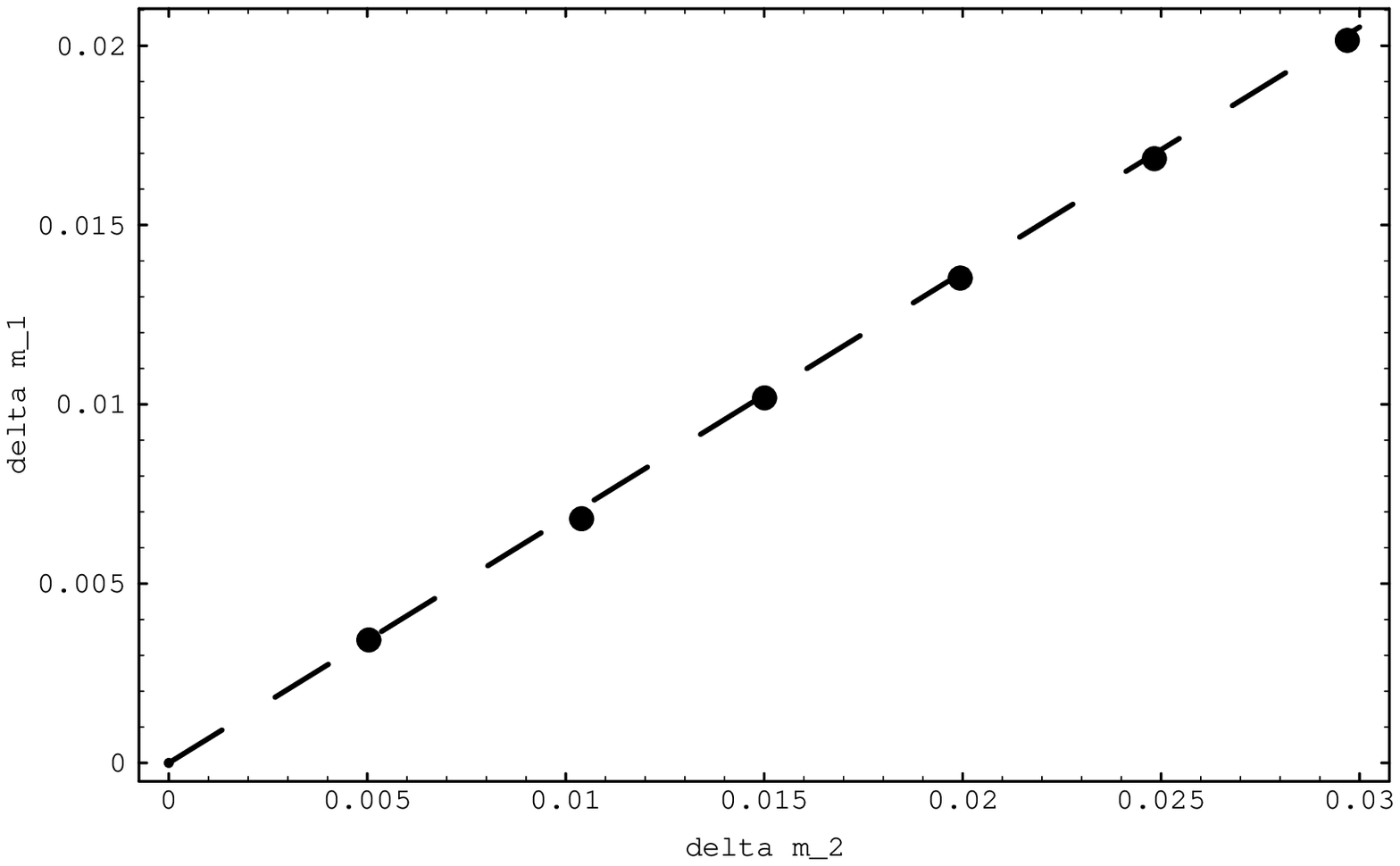}}
\vspace{-5cm}
\begin{center}
{\bf Figure 4}
\end{center}
\end{figure}

\begin{figure}[h]
\centerline{\hspace{2cm}\psfig{figure=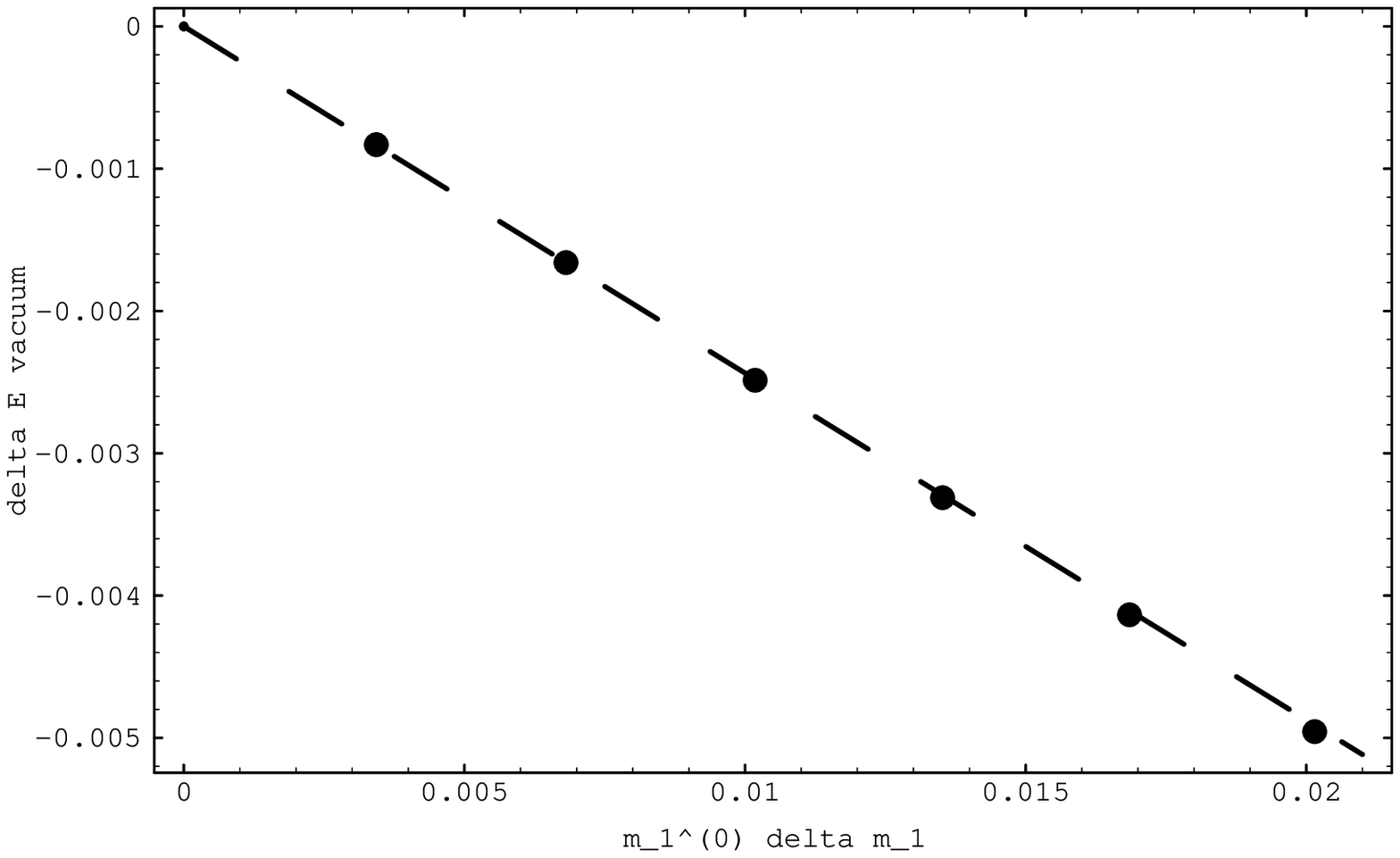}}
\vspace{-5cm}
\begin{center}
{\bf Figure 5}
\end{center}
\end{figure}

\end{document}